\documentclass[3p,times,twocolumn]{elsarticle}

\usepackage{ecrc}

\volume{00}

\firstpage{1}

\journalname{Nuclear Physics B Proceedings Supplement}

\runauth{S.G. Bondarenko et al.}

\jid{nuphbp}

\jnltitlelogo{Nuclear Physics B Proceedings Supplement}

\usepackage{amssymb}
\def\tlab{\mbox{$T_{\rm Lab}$}}

\def\lamr{\lambda^r}
\def\lami{\lambda^i}

\def\pzp{p_0^{\prime}}

\newcommand{\bp}{{\bf p}}
\newcommand{\bpp}{{\bf p}^{\prime}}
\newcommand{\bk}{{\bf k}}

\def\mth{m_{th}}

\begin{document}

\begin{frontmatter}



\dochead{}

\title{Relativistic complex separable potential for describing the neutron-proton system \\
in $^3S_1$-$^3D_1$ partial-wave state}


\author{S.G. Bondarenko, V.V. Burov, E.P. Rogochaya}

\address{JINR - Joint Institute for Nuclear Research, Joliot-Curie 6, 141980 Dubna, Moscow region, Russia}

\begin{abstract}
Within a covariant Bethe-Salpeter approach the relativistic
complex separable kernel of the neutron-proton interaction for the
coupled $^3S_1^+$-$^3D_1^+$ partial-wave state is constructed.
The rank-six separable potential elaborated earlier is real-valued,
and therefore makes it possible to describe only the elastic
part (phase shifts, low-energy parameters, deuteron properties, etc.) of the elastic
neutron-proton scattering. The description of the inelasticity
parameter comes out of the imaginary part introduced intthe
potential. The complex potential parameters are obtained using
the available elastic neutron-proton scattering experimental data up to 1.1\,GeV.
\end{abstract}

\begin{keyword}
elastic neutron-proton scattering \sep Bethe-Salpeter equation \sep partial-wave analysis
\sep separable kernel \sep inelasticity



\end{keyword}

\end{frontmatter}

% \linenumbers

\section{Introduction}
\label{intro} A theoretical description of the neutron-proton
($np$) system makes it possible to understand the nuclear
structure. The nucleon-nucleon (NN) scattering is
traditionally described in terms of the meson-nucleon theory at
low and medium energies. The one-boson exchange
(realistic) potential plays a predominant role in this case. At higher
energies, production processes and inelasticities become important,
and systems composed of nucleons and mesons contribute to the NN
scattering.  At present there is no quantitative description of
the NN scattering above the inelastic threshold neither in terms of
QCD nor of meson and nucleon degrees of freedom.

There have been several theoretical attempts built within boson
exchange models to explain NN scattering data up to 1\,GeV. All of
them describe observables perfectly for energies up to 300\,MeV.
However, using boson exchange potentials at higher energies gives
only a qualitative agreement with experimental data.
Optical model studies have been suggested for the medium and
high energy NN scattering
\cite{Khokhlov:2005vp,Funk:2001ph}. They give a quantitative
agreement with the experimental data for the NN scattering and
Bremsstrahlung within the wide energy range.

The Bethe-Salpeter (BS)
approach~\cite{Salpeter:1951sz,Bondarenko:2002zz},
as a four-dimensional covariant formalism, is the most
appropriate to study the NN interaction. It is convenient to use a separable
ansatz~\cite{Bondarenko:2002zz} for the interaction kernel in the
BS equation in this picture. Papers \cite{npa1,npa2} have
suggested multirank separable potentials to describe the $np$
scattering with the total angular momentum $J=0,1$ and the
deuteron. The obtained potentials can describe all available
experimental data for the phase shifts, static properties of the
deuteron, and the exclusive electron-deuteron breakup in the
plane-wave approximation~\cite{npa1,npa2,fbs}.

However, it is important to investigate the influence of the
inelastisity in the elastic NN scattering.
We describe the inelasticity using a complex NN potential of a
special type \cite{Bondarenko:2011hs} instead of the real-valued
one obtained earlier in Refs.~\cite{npa1,npa2}. The method has been
successfully used for the description of the uncoupled
partial-wave states of the $np$ system with the total angular
momentum $J=0,1$. Now we apply it to construct the complex
potential for the coupled $^3S_1^+$-$^3D_1^+$ partial-wave state.

The paper is organized as follows. In Sec.\ref{sect2}, the
parametrization of the scattering matrix is described. The used
complex separable interaction kernel is considered in
Sec.\ref{sect3}. The procedure which we apply to find new
imaginary interaction kernel parameters and the obtained
parameters are presented in Sec.\ref{sect4}. The discussion and
conclusion are given in Sec.\ref{sect5}.

\section{Parametrization of the S matrix}\label{sect2}
Following Arndt et al.~\cite{arndtroper}, we prefer to use the
parametrization of the $K$ matrix rather than the scattering
matrix $S$:
\begin{eqnarray}
K = i\frac{1-S}{1+S}={\rm Re}K+i{\rm Im}K,
\end{eqnarray}
where $K$ is a 2$\times$2 matrix for the coupled partial-wave state.
The real part ${\rm Re}K$ is represented as
\begin{eqnarray}
{\rm Re}K = i\frac{1-S_e}{1+S_e},
\end{eqnarray}
where
\begin{eqnarray}
S_e=\left(
\begin{array}{cc}
\cos2\varepsilon_1 e^{2i\delta_<} & i\sin2\varepsilon_1 e^{i(\delta_<+\delta_>)} \\
i\sin2\varepsilon_1 e^{i(\delta_<+\delta_>)} & \cos2\varepsilon_1
e^{2i\delta_>}
\end{array}
\right)
\end{eqnarray}
is the well-known Stapp parametrization \cite{Stapp:1956mz} for
the elastic NN scattering matrix. Here $\delta_<=\delta_{L=J-1}$,
$\delta_{>}=\delta_{L=J+1}$ are phase shifts of $^3S_1^+$ and
$^3D_1^+$ states, respectively, and $\varepsilon_1$ is a mixing
parameter. The ${\rm Im}K$ is given by
\begin{eqnarray}
{\rm Im}K=&&\\
&&\hspace*{-7mm}\left(
\begin{array}{cc}
\tan^2\rho_< & \tan\rho_<\tan\rho_>\cos\mu \\
\tan\rho_<\tan\rho_>\cos\mu & \tan^2\rho_>
\end{array}
\right)\nonumber
\end{eqnarray}
in terms of the inelasticity parameters $\rho_{<,>}, \mu$ corresponding
to the states with the orbital momenta $L=J-1,J+1$, and $\varepsilon_1$ respectively.
%It should be noted that we have calculated parameter $\mu$ but not present it below
%due to absence of the experimental data~\cite{said}.
%
\section{Complex separable kernel}\label{sect3}
We assume that the interaction kernel $V$ conserves parity, the
total angular momentum  $J$ and its projection, and isotopic spin.
Due to the tensor nuclear force, the orbital angular momentum $L$
is not conserved. The negative-energy two-nucleon states are
switched off, what leads to the total spin $S$ conservation. The
partial-wave-decomposed BS equation in the center-of-mass system
of the $np$ pair is therefore reduced to the following form:
\vskip 5mm
\begin{eqnarray}
&&\hspace*{-7mm}T_{l'l}(\pzp, |\bpp|; p_0, |\bp|; s) =
\label{BS} \\
&&\hspace*{-7mm} V_{l'l}(\pzp, |\bpp|; p_0, |\bp|; s)+
\frac{i}{4\pi^3}\sum_{l''}
\int\limits_{-\infty}^{+\infty}\! dk_0\int\limits_0^\infty\! \bk^2 d|\bk|\,  \times \nonumber\\
&&\hspace*{-7mm}\frac{V_{l'l''}(\pzp, |\bpp|; k_0,|\bk|; s)\,
T_{l''l}(k_0,|\bk|;p_0,|\bp|;s)}
{(\sqrt{s}/2-E_{\bk}+i\epsilon)^2-k_0^2}.\nonumber
\end{eqnarray}
%where $l=l^{\prime}=l^{\prime\prime}$ for spin-singlet and
%uncoupled spin-triplet states.
The square of the $np$ pair total
momentum $s$ is related with the laboratory energy $\tlab$ as:
$s=2m\tlab+4m^2$, $m$ is the mass of the nucleon.

To describe the inelasticity in the elastic NN scattering, we
modify the real-valued relativistic potential adding the imaginary
part:
$$ V_r \to V = V_r + iV_i.$$

To solve the Eq.(\ref{BS}), the separable (rank $N$)
ansatz~\cite{Bondarenko:2002zz} for the NN interaction kernel is
used:
\begin{eqnarray}
&&\hspace*{-7mm}V_{l'l}(\pzp, |\bpp|; p_0, |\bp|; s)= \label{V_separ}\\
&&\hspace*{-7mm}\sum_{m,n=1}^N \Big[\lamr_{mn}(s) + i
\lami_{mn}(s)\Big] g_i^{[l']}(\pzp, |\bpp|)g_j^{[l]}(p_0,
|\bp|),\nonumber
\end{eqnarray}
where the imaginary part $\lami$ has the following form~\cite{Bondarenko:2011hs}:
\begin{eqnarray}
\lami_{mn}(s) = \theta(s-s_{th})\, \Big(1-\frac{s_{th}}{s}\Big)\,{\bar\lambda}^i_{mn},
\label{lami}
\end{eqnarray}
$g_j^{[l]}$ are model functions, $\lambda_{mn} = \lambda^{r}_{mn}+i\lambda^{i}_{mn}$
is a matrix of model parameters and $s_{th}$ is
an inelasticity threshold value (the first energy point where the inelasticity
becomes nonzero). In this case, the resulting $T$ matrix has a similar
separable form:
\begin{eqnarray}
&&T_{l'l}(\pzp, |\bpp|; p_0, |\bp|; s)= \\
&&\sum_{m,n=1}^N\tau_{mn}(s)g_i^{[l']}(\pzp, |\bpp|)
g_j^{[l]}(p_0, |\bp|),\nonumber
\end{eqnarray}
where
\begin{eqnarray}
\big(\tau_{mn}(s)\big)^{-1} = \Big(\lamr_{mn}(s)+i\lami_{mn}(s)\Big)^{-1}+h_{mn}(s),
\end{eqnarray}
\begin{eqnarray}
&&\hspace*{-9mm}h_{mn}(s)= \label{H_separ}\\
&&\hspace*{-9mm}-\frac{i}{4\pi^3}\sum_{l}\int dk_0\int
\bk^2d|\bk|\frac{g_m^{[l]}(k_0,|\bk|)g_n^{[l]}(k_0,|\bk|)}{(\sqrt
s/2-E_{\bk}+i\epsilon)^2-k_0^2}.\nonumber
\end{eqnarray}
The functions $g_m^{[l]}$ and the parameters $\lamr$ coincide with those
used in Ref.~\cite{npa2} while $\lami$ are new parameters which are
calculated.

Thus, we introduce the imaginary part $V_i$ of the potential $V$
(\ref{V_separ}) adding the new parameters $\lami$ to the real part
$V_r$ which is left unchanged.  It allows us to describe the
additional inelasticity parameters by a minimal change of the
previous kernel~\cite{npa2}.
\section{Calculations and results}\label{sect4}
The description of the inelastic part in the $^3S_1^+$-$^3D_1^+$
state is performed in the same way as in our previous paper
\cite{Bondarenko:2011hs} for the uncoupled partial-wave states. We
start from the real-valued rank-six interaction kernel MY6
\cite{npa2} obtained from the description of low-energy
characteristics and phase shifts for the laboratory energies
$\tlab$ up to 1.1\,GeV whose experimental values were taken from the
SAID program \cite{said}. The parameters of the real part are
fixed. Then $\lami$ are calculated to give a correct behavior of
the inelasticity parameters $\rho_<$, $\rho_>$. The experimental
data for them can be taken from the SAID program.

The minimization procedure for the function
\begin{eqnarray}
\chi^2=&&\hspace*{-8mm}
\sum\limits_{m=\mth}^{n}\sum\limits_{l=<,>}
(\delta^{\rm exp}_l(s_m)-\delta_l(s_m))^2/(\Delta\delta^{\rm exp}_l(s_m))^2
\nonumber\\ &&\hspace*{2mm}+
%\sum\limits_{m=\mth}^{n}
(\rho^{\rm exp}_l(s_m)-\rho_l(s_m))^2/(\Delta\rho^{\rm exp}_l(s_m))^2
\label{mini_p}
\end{eqnarray}
is used. Here $n$ is the number of available experimental points,
$\mth$ is the number of the data point corresponding to the
first nonzero $\rho$ value. It is defined by the threshold kinetic
energy $\tlab_{\,th}$ which is taken from the single-energy
analysis~\cite{said}.

The resulting parameters $\lami$ are listed in
Table~\ref{lambdasi}\footnote{We would like to note misprints in
Tables 2 and 3~\cite{npa2} where $\bar\lambda$ (GeV$^2$) should be
read as $\bar\lambda$ (GeV$^4$).}. In Figs.\ref{3s1}-\ref{eps},
the results of the phase shift, inelasticity parameter and mixing
parameter calculations (MYI6 - red dashed line) are compared with
the experimental data, our previous result without
inelasticities~\cite{npa2} (MY6 - red solid line; only for the
phase shifts and the mixing parameter), the SP07
solution~\cite{Arndt:2007qn} (green dashed-dotted line) and the
optical potential FGA \cite{Funk:2001ph} (blue
dashed-dotted-dotted line).
\begin{center}
\begin{table}[h]
\caption{Parameters $\bar\lami$ of the rank-six kernel MYI6 for
the coupled $^3S_1^+$-$^3D_1^+$ state. } \centering
\begin{tabular}{lcclc}
%\hline\hline
%
%                               &  MYI6          \\
\hline\hline\\[-4mm]
$\bar\lami_{11}$    (GeV$^2$) &  -366.6091   \\
$\bar\lami_{12}$    (GeV$^2$) &   128.6970   \\
$\bar\lami_{13}$    (GeV$^2$) &  -0.3089566  \\
$\bar\lami_{14}$    (GeV$^2$) &  -68.07828   \\
$\bar\lami_{15}$    (GeV$^2$) &   64.67047   \\
$\bar\lami_{16}$    (GeV$^2$) &  -11.51058   \\
$\bar\lami_{22}$    (GeV$^2$) &   68.53852   \\
$\bar\lami_{23}$    (GeV$^2$) &   3.129745   \\
$\bar\lami_{24}$    (GeV$^2$) &   458.8815   \\
$\bar\lami_{25}$    (GeV$^2$) &   2669.315   \\
$\bar\lami_{26}$    (GeV$^2$) &   11.35009   \\
$\bar\lami_{33}$    (GeV$^2$) &  -2.395499   \\
$\bar\lami_{34}$    (GeV$^2$) &   15.130951  \\
$\bar\lami_{35}$    (GeV$^2$) &  -55.78875   \\
$\bar\lami_{36}$    (GeV$^2$) &   0.4364534  \\
$\bar\lami_{44}$    (GeV$^2$) &   411.3334   \\
$\bar\lami_{45}$    (GeV$^2$) &   3280.671   \\
$\bar\lami_{46}$    (GeV$^2$) &  -11.29448   \\
$\bar\lami_{55}$    (GeV$^2$) &   5685.375   \\
$\bar\lami_{56}$    (GeV$^2$) &  -44.92390   \\
$\bar\lami_{66}$    (GeV$^2$) &  -0.07438475 \\
$\tlab_{\,th}$      (GeV$$)   &   0.4      \\
\hline\hline
\end{tabular}\label{lambdasi}
\end{table}
\end{center}
\section{Discussion and conclusion}\label{sect5}
\begin{figure}[h]
\begin{center}
\includegraphics[width=0.49\textwidth]{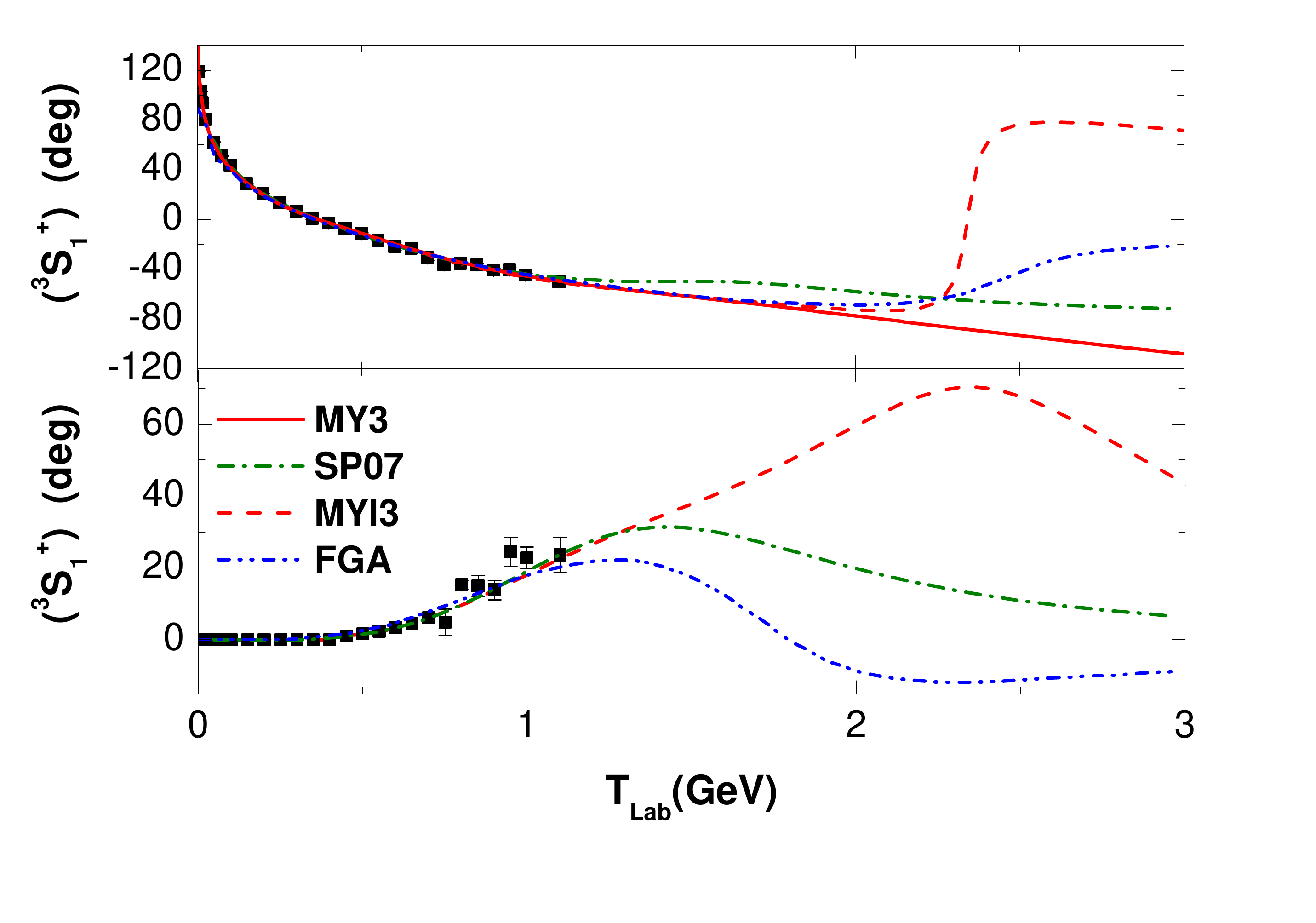}
\caption{{Phase shifts and inelasticity parameter for the $^3S_1^+$ partial-wave state. The results of calculations with the potentials - real-valued MY6 \cite{npa2}, complex MYI6, optical FGA \cite{Funk:2001ph} and the SP07 solution \cite{Arndt:2007qn} are compared.}}
\label{3s1}
\end{center}
\end{figure}

The obtained phase shifts and inelasticity parameter for the
$^3S_1^+$ partial-wave state are presented in Fig.\ref{3s1}. We
see that all considered calculations (MY6, MYI6, SP07, FGA) give an
excellent description of all available experimental data for phase
shifts and are quite similar to each other for $\tlab$ up to
2\,GeV. However, MYI6 model phase shifts suddenly change their
behavior at the laboratory energy $\tlab\gtrsim$ 2\,GeV, which
becomes similar to FGA (see also SM97, FA00 and SP00 solutions
\cite{Arndt:2000xc}). Therefore, to understand whether this
behavior means some physical effect or the model needs to be
improved, the experimental data at $\tlab>$1.1\,GeV are
necessary. The descrpition of the inelasticity parameter is
perfect for the MYI6, FGA potentials and the SP07 solution for $\tlab$
up to 1.1\,GeV where the experimental data are available and they
are different at higher energies.

\begin{figure}[h]
\begin{center}
\includegraphics[width=0.49\textwidth]{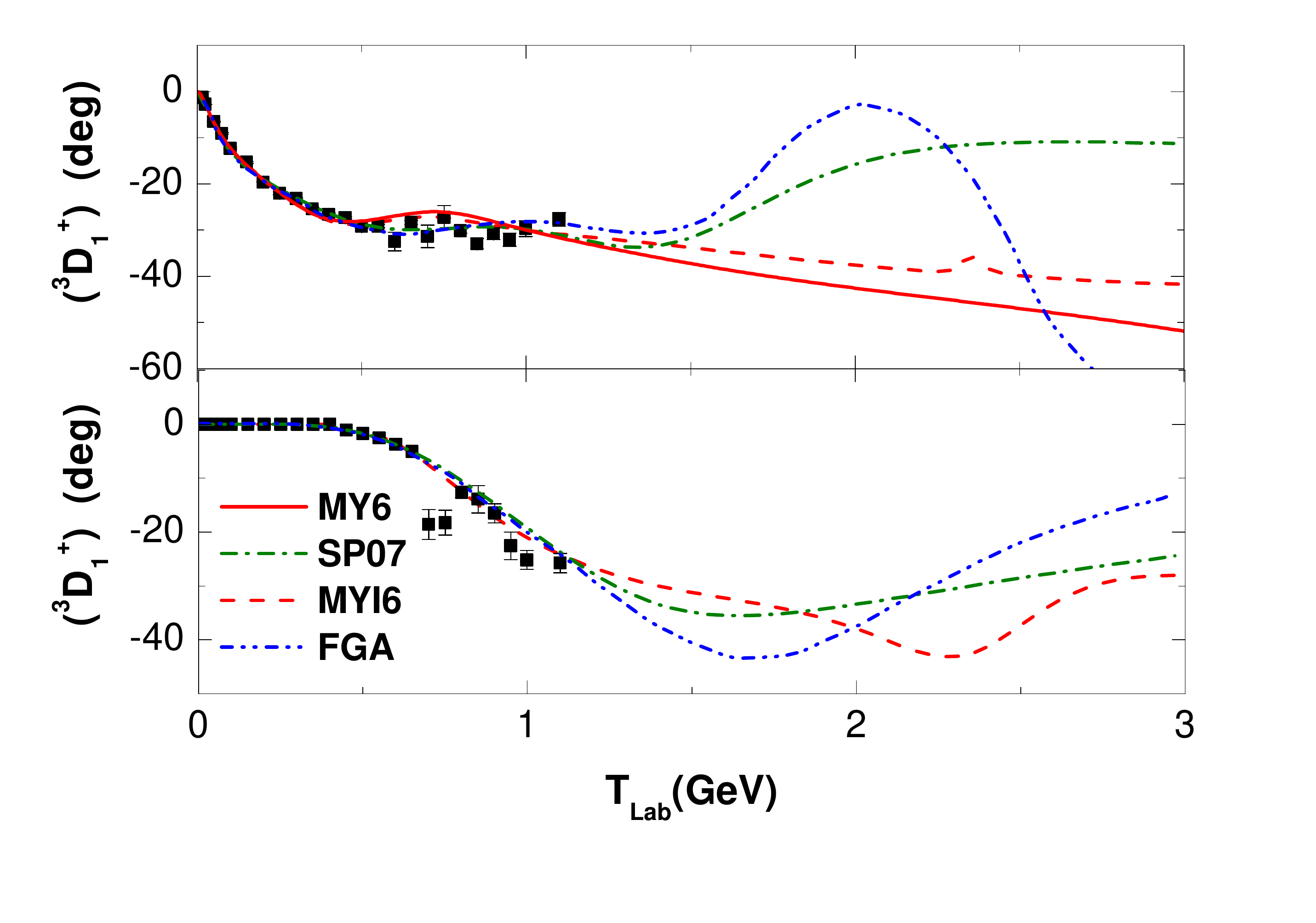}
\caption{{Phase shifts and inelasticity parameter for the $^3D_1^+$ partial-wave state. The notations are the same as in Fig.\ref{3s1}.}}
\label{3d1}
\end{center}
\end{figure}

Fig.\ref{3d1} shows the results of the calculations for the
$^3D_1^+$ state. All of model calculations (MY6, MYI6, SP07, FGA)
are in a good agreement with the experimental data and
different at the laboratory energies $\tlab\gtrsim$ 1.1\,GeV.

\begin{figure}[h]
\begin{center}
\includegraphics[width=0.49\textwidth]{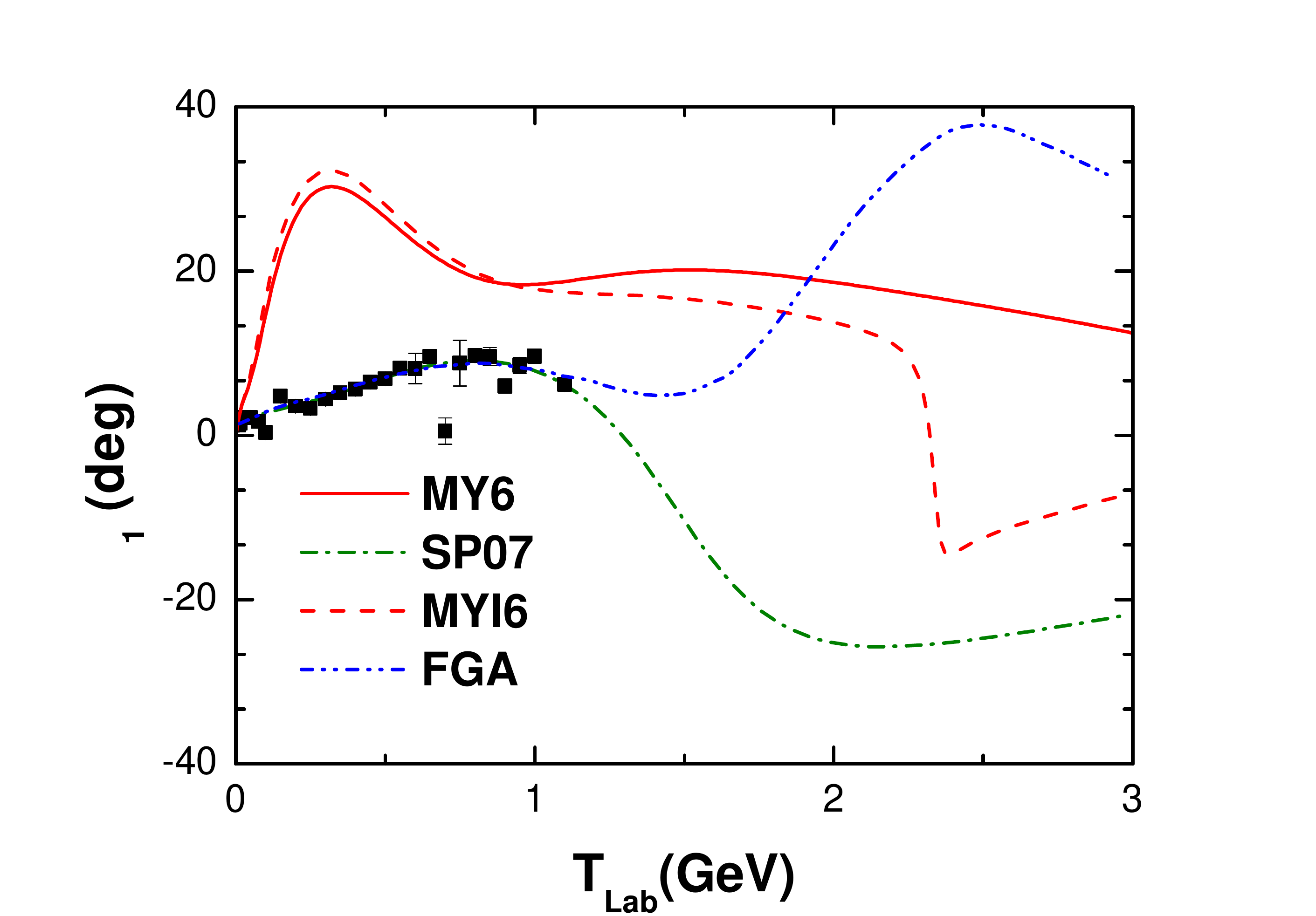}
\caption{{Mixing parameter $\varepsilon_1$. The notations are the same as in Fig.\ref{3s1}.}}
\label{eps}
\end{center}
\end{figure}

The mixing parameter $\varepsilon_1$ is depicted in Fig.\ref{eps}.
As it is seen from the figure, the MY6 model does not
describe it at all. The MYI6 model has similar behavior as MY6
and differ from it at the kinetic energies $\tlab\gtrsim$ 1.1\,GeV.

It is seen that the proposed MYI6 potential gives a consistent
description of the existing experimental data for the phase shifts
and the inelasticity parameter in the coupled $^3S_1^+$-$^3D_1^+$
partial-wave state. It should be noted that since all parameters
of the real-valued separable interaction kernel ($\lamr$, $\beta$ and
$\alpha$) found in the previous analysis~\cite{npa2} have been
fixed, the MY6 and MYI6 model phase shifts
coincide up to $\tlab < \tlab_{\,th}$ and are different
at $\tlab> \tlab_{\,th}$. The comparison of four different
calculations MY6, MYI6, SP07, FGA has shown that the experimental data
for the phase shifts and inelasticity parameter at higher
laboratory energies are vitaly necessary to define the behavior of the NN
potential precisely.
\section{Acknowledgments}\label{sect6}
We would like to thank the organizers of the 5th Joint International
HADRON STRUCTURE'11 Conference (June 27th - July 1st, 2011, Tatransk$\acute{\rm a}$ $\breve{\rm S}$trba,
Slovak Republic) for invitation, support and opportunity to present our new results.
%



%% References

%\nocite{*}
%\bibliographystyle{elsarticle-num}
%\bibliography{martin}



\end{document}